\documentclass[useAMS,usenatbib,preprint]{mn2e}
\usepackage{graphicx}
\usepackage{color}
\usepackage{aas_macros}
\usepackage{gensymb}
\usepackage[T1]{fontenc}
\usepackage{float}
\title[High redshift radio galaxies and divergence from the CMB dipole]
{High redshift radio galaxies and divergence from the CMB dipole }

\author[J. Colin, R. Mohayaee, M. Rameez \& S. Sarkar]
{
Jacques Colin$^{1,2}$,
Roya Mohayaee$^{1,2}$,
Mohamed Rameez$^2$,
Subir Sarkar$^{2,3}$
\\
$^{1}$ Sorbonne Universit\'es, UPMC Univ Paris 06, CNRS, Institut d'Astrophysique de Paris, 98bis Bld Arago, Paris 75014, France\\
$^{2}$ Niels Bohr Institute, University of Copenhagen, Blegdamsvej 17, 2100 Copenhagen {\O}, Denmark\\
$^{3}$ Rudolf Peierls Centre for Theoretical Physics, University of Oxford, 1 Keble Road, Oxford, OX1 3NP, United Kingdom
}

\date{\today}

\begin{document}
\maketitle
\label{firstpage}

\begin{abstract}
Previous studies have found our velocity in the rest frame of radio galaxies at high redshift to be much larger than that inferred from the  dipole anisotropy of the cosmic microwave background. We construct a full sky catalogue, NVSUMSS, by merging the NVSS and SUMSS catalogues and removing local sources by various means including cross-correlating with the 2MRS catalogue. We take into account both aberration and Doppler boost to deduce our velocity from the hemispheric number count asymmetry, as well as via a 3-dimensional linear estimator. Both its magnitude and direction depend on cuts made to the catalogue, e.g. on the lowest source flux, however these effects are small. From the hemispheric number count asymmetry we obtain a velocity of $1729 \pm 187$ km\,s$^{-1}$ i.e. about 4 times \emph{larger} than that obtained from the CMB dipole, but close in direction, towards RA$=149\degree \pm 2\degree$, dec = $-17\degree \pm 12\degree$. With the 3-dimensional estimator, the derived velocity is $1355 \pm 174$ km\,s$^{-1}$ towards RA$=141\degree \pm 11\degree$, dec=$-9\degree \pm 10\degree$. We assess the statistical significance of these results by comparison with catalogues of random distributions, finding it to be $2.81 \sigma$ (99.75\% confidence).
\end{abstract}


\section{Introduction}
\label{sec:introduction}

The cosmic microwave background (CMB) exhibits a dipole anisotropy \citep{bracewell68,henry71,smoot77} which is a thousand times bigger than the primordial temperature fluctuations first detected by COBE \citep{kogut93}. An observer moving through an isotropic radiation field would observe such a dipole \citep{stewart67,peebles68} and conversely the observed dipole can be used to infer the velocity of the observer. Indeed the velocity of the Solar system (and our Local Group) is inferred from the CMB temperature dipole assuming that the latter is entirely due to our motion. 
 
Although a kinematic origin is the most plausible interpretation of the CMB dipole, one can consider other mechanisms which may be, at least partially, responsible \citep{paczynski90,langlois96}. An independent measurement of our velocity is therefore desirable. It has been suggested that this can be done by measuring the aberration of the CMB \citep{challinor02,burles06}, however the effect is too small to be detected with $>3\sigma$ significance even using the best available data to date \citep{planck14}.

If the dipole is due to our motion then it requires the Universe to be anisotropic at least locally. Indeed galaxy surveys do exhibit such an anisotropy. Along the direction of the CMB dipole lie the most massive neighbouring superclusters: Virgo, the Great Attractor Hydra-Centaurus, Coma, Hercules and Shapley, and possibly yet-to-be-mapped superclusters beyond these (see Table \ref{tab:localsuperclusters}). The gravitational attraction of these structures is evidently pulling us in the direction indicated by the CMB dipole. However detailed maps of the local Universe, {\it e.g.} from the infrared 2MASS Redshift Survey (2MRS) or supernova catalogues, show that the full dipole \emph{cannot} be attributed to these structures and at least 20\% remains unaccounted for \citep{lauer94,hudson04,watkins09,lavaux10,feldman10,colin11,6df12,feindt13,watkins15}. It is thus necessary to go beyond even Shapley (at $\sim262$~Mpc) to establish if there is indeed convergence to the CMB frame at larger scales --- as is essential if the universe is to be described as homogeneous when averaged on such scales. 

However, there are no full-sky optical galaxy redshift surveys extending well beyond Shapley and previous studies in this context have therefore suffered from lack of full sky coverage ({\it e.g.} \cite{coles93,itoh10}). Further complications arise when using other catalogues such as of galaxy clusters to search for the expected dipole signal ({\it e.g.} \citet{chluba05}). There have in fact been persistent reports of an anomalous `dark flow' on large scales traced by the kinematic Sunyaev-Zel'dovich effect in clusters, in roughly the same direction as the CMB dipole \citep{Kashlinsky08,Atrio-Barandela15}.

On the other hand, radio galaxies are extremely luminous, can be observed up to high redshifts, and are unaffected by dust obscuration. However early analyses of the dipole in the distribution of radio galaxies suffered from the severe sparseness of their catalogues \citep{lahav98}. Hence such studies have become feasible only after the NRAO VLA Sky Survey (NVSS). This is a catalogue of nearly 2 million radio galaxies covering the sky above declination of $-40\degree$. NVSS has been used extensively to search for convergence to the CMB dipole and most previous studies have found an interesting \emph{discrepancy} with the CMB measurement of our local velocity \citep{singal11,gibelyou12,rubart13,tiwari15,tiwarijain15}. 

While the velocity of the Solar system barycentre inferred from the CMB temperature dipole anisotropy is 369 km~s$^{-1}$, the value inferred from NVSS ranges from 700 km~s$^{-1}$ \citep{blake02} to over 2000 km~s$^{-1}$ \citep{gibelyou12}. We too find a similarly large velocity and demonstrate that it depends only mildly on the characteristics of the datasets used: {\it e.g.} the width of the Galactic and super-Galactic plane strips removed from the survey and the lower and upper limits on the flux. Significantly the direction of the velocity is quite robust and well-aligned with the direction of the CMB dipole. 

It is hard to understand how such a large difference can have a physical explanation. Previous works \citep{paczynski90,langlois96,ghosh14} have invoked horizon-scale isocurvature perturbations as an alternative to the kinematic origin for the CMB dipole. Here however, we are observing a dipole far \emph{larger} than that inferred from CMB itself, but still at high redshift. In addition local data --- specifically distance and redshift catalogues --- indicate that at low redshifts a good part of the CMB dipole is recovered by $\sim250$~Mpc. Although the present data does not allow us to fully explore how the convergence to CMB develops as we go deeper into the Universe and in  redshift space, the radio data at redshift $z \sim 1$ clearly indicates that a significantly \emph{larger} dipole exists in the rest frame of the radio galaxies. The origin of this difference from the CMB dipole is a real mystery and deserves closer examination.

A possible explanation is systematics due to the incomplete sky coverage of NVSS which can generate a spurious dipole \citep{gibelyou12,tiwari16}. Various methods have been used to restore the symmetry required for a robust dipole analysis such as cutting out the NVSS catalogue above declination of $40\degree$ \citep{singal11} or removing the Galactic and counter-Galactic plane \citep{rubart13}. Our work elaborates on previous studies in several ways. We construct a \emph{full-sky} catalogue, which we call NVSUMSS, by patching the NVSS with its southern-sky counterpart --- the Sydney University Molonglo Sky Survey (SUMSS). This results in a catalogue with about 600,000 radio sources with flux above 10 mJy after cutting out a strip of $\pm 10\degree$ around the Galactic plane in NVSS to match the corresponding cut in SUMSS.

Another possible systematic which has often been discussed is contamination by local sources. It is well known that a large part of the CMB dipole anisotropy is due to the anisotropic local distribution of matter \citep{lavaux10,colin11}. Past studies have tried to clean NVSS from such local source contamination either by removing individual large local structures \citep{blake02} or by cutting out the super-Galactic plane in which the majority of local superclusters lie \citep{tiwari15}. However even after all these cuts the radio dipole extracted by all previous studies remains well above that inferred from the CMB. To fully eliminate the contribution from the local structures, we adopt several strategies. First we cross-correlate NVSUMSS with the catalogue \citep{falcke12} of local radio sources (LRS) and remove patches (of increasing area) around these sources from NVSUMSS catalogue. Our results are not however significantly altered by doing so. As the LRS catalogue is limited to sources brighter than 200 mJy, we go further by next removing eight dominant local superclusters from Virgo to Ophiuchus. Previous studies have shown that these local structures are  responsible for nearly $80\%$ of the CMB dipole. Even so their removal does not alter our results significantly. We then remove the super-Galactic plane completely by cutting out a $\pm 10\degree$ strip around it. This too has little effect. Finally we cross-correlate NVSUMSS with the 2MRS catalogue, and remove zones of increasingly larger areas around the 2MRS sources from the NVSUMSS catalogue. This too has no significant impact.

Because of evolution, the fraction of nearby radio sources is low even at large flux densities, the median redshift of radio sources in complete samples with flux ranging from mJy to Jy being $z \sim 0.8$ \citep{nvss89}. The fact that the redshift distribution of complete samples of radio sources peaks at $z \sim 1$ for flux-densities down to $\sim 10$ mJy implies that local sources are swamped by more numerous distant ones \citep{dezotti10}. The mean redshift of NVSS sources is estimated to be above 1 \citep{ho08,dezotti10}. Thus most of the sources in our NVSUMSS catalogue lie at high redshift and by removing local sources through cross-correlation of NVSUMSS with the 2MRS catalogue we ensure that they constitute an uniform background. Our motion with respect to such an uniform background causes the aberration of light and the sources would thus be displaced in the sky by an angle that depends on our velocity. More sources would be observed in the direction of the motion than in the opposite direction. 

Thus by measuring the aberration angle, we infer the velocity of the Solar system (or the Local Group) barycentre in the rest frame of the distant radio galaxies. A catalogue with good sky coverage would exhibit a hemispherical asymmetry which peaks when the plane of hemisphere is perpendicular to the direction of the motion. A simple formula can be used to determine  the aberration angle and hence our velocity in the rest frame of the radio sources. Aberration analysis is very powerful in the sense that it is \emph{independent} of distance so can be used in conjunction with a full-sky photometric galaxy survey, as long as there are no observational biases generating spurious anisotropies. In fact the motion of the observer also results in Doppler boosting of the frequency at which the sources are observed --- as a result more sources will be observed in the direction of the motion. Unlike aberration, Doppler boosting depends on the flux and it is known that for radio sources the two effects affect the number-count almost equally. 

This paper is organised as follows: in Section \ref{sec:method} we describe our method, and search for dipole anisotropy in the NVSS data alone (Section \ref{sec:nvss}). Then we introduce the SUMSS catalogue (Section \ref{sec:sumss}) and  patch NVSS and SUMSS together (Section \ref{sec:nvsumss}), exploring different ways of doing so. In Section \ref{sec:local} we perform a thorough clean up of local sources by cross-correlating NVSUMSS with the 2MRS catalogue. We investigate the effects of various cuts made on the inferred dipole velocity and discuss the statistical significance of the result (Section \ref{sec:ss}). We present our conclusions in Section \ref{sec:conclusion}.


 \begin{table}
\caption{Local superclusters which are responsible for a large part of the CMB dipole \citep{lavaux10,colin11}.} 
\begin{tabular}{|c|c|c|}
& & \\
\hline
 Name & $\{ {\rm RA \degree, dec \degree} \}$ \\
 \hline
 \hline
 Virgo &$\{186.8 , 12\}$ \\
 \hline
 Hydra \& Centaurus & $\{192.2 , -38\}$ \\
 (Great Attractor) & \\
 \hline
 Norma & $\{243 , -59 \} $ \\
 \hline
 Persus-Pisces & $\{27.5 , 36 \}$ \\
\hline
 Coma & $\{195.0 , 28.0 \}$ \\
 \hline
 Hercules & $\{241.3 , 17.8 \}$ \\
 \hline
 Shapley & $\{201.3 , -30 \}$ \\
 \hline
 Ophiuchus & $\{255 , -8 \}$ \\
 \label{tab:localsuperclusters}
 \end{tabular}
\end{table}

\section{The method}
\label{sec:method}

The aberration angle, $\Delta\theta$, by which a galaxy is displaced due to the motion of an observer moving with velocity $v$ is ({\it e.g.} \citet{kopeikin11}) for small velocities $v \ll c$:
\begin{equation}
 \Delta\theta = (|v|/c)\ {\rm sin}\theta .
 \label{eq:aberration}
 \end{equation}
To evaluate this we simply count the number of galaxies in hemispheres and use the relation of the hemispherical number count asymmetry, $\delta N$, to the aberration angle:
 \begin{equation}
 \Delta\theta=2(\delta N/N) .
 \end{equation}
Here we have assumed a homogeneous distribution with an average number surface density, and not taken into account any masks and cuts on the catalogue. 

The total number asymmetry is due to both aberration and Doppler boosting. The expected amplitude of the latter is \citep{baldwin84}: $(v/c)[2+x(1+ \alpha)]$, where $x$ and $\alpha$ are flux indices, defined through the integral source counts
\begin{equation}
{\rm d}N/{\rm d}\Omega (>S) = k S^{-x} ,
\label{eq:ellisbaldwin}
\end{equation}
and the flux density at a fixed observing frequency
\begin{equation}
S_{\rm obs}=S_{\rm rest}\delta^{1+\alpha} . 
\end{equation}
It is straightforward to evaluate $x$ for a given catalogue, e.g. as shown in Fig.~\ref{fig:nvss} it varies from 0.894  at low flux to 1.68 at high flux for NVSS. The Doppler shift is more important for faint galaxies, hence the value at low flux is  most relevant. The value of $\alpha$ is usually taken to be $0.75$ \citep{baldwin84}. Hence the aberration and Doppler boosting contribute almost equally to the number count asymmetry --- 53\% and 47\% respectively. 
 
As we are looking specifically for a dipole anisotropy (to compare with the CMB dipole), we do not use spherical harmonic decomposition. The simplest and most robust estimate of the dipole is just the hemispheric asymmetry in the number count of sources. We calculate this as follows: first by  randomly selecting a direction and counting the number of galaxies in that hemisphere. Secondly by regularly rotating a hemisphere centred on zero declination and looking for the right ascension that gives the maximum asymmetry. Third by scanning the sky using directions from a HEALPix map \citep{healpix04} with Nside = 32 (12,288 pixels) which are not random directions but regularly spaced. As we shall show, all these methods give similar results in agreement with each other. While the hemispheric asymmetry estimator is robust and allows the estimation of the variability of the dipole around its directional maximum, its statistical variability is high. Hence, we also evaluate the dipole using a 3-dimensional linear estimator \citep{crawford09}. Other estimators have been proposed in the literature \citep{chenschwarz2016}.

\section{NRAO VLA Sky Survey (NVSS)}
\label{sec:nvss}

\begin{figure}
\includegraphics[width=0.45\columnwidth]{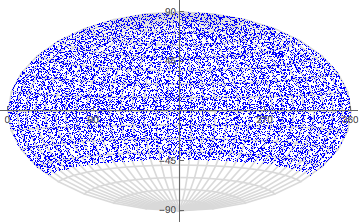} 
\includegraphics[width=0.45\columnwidth]{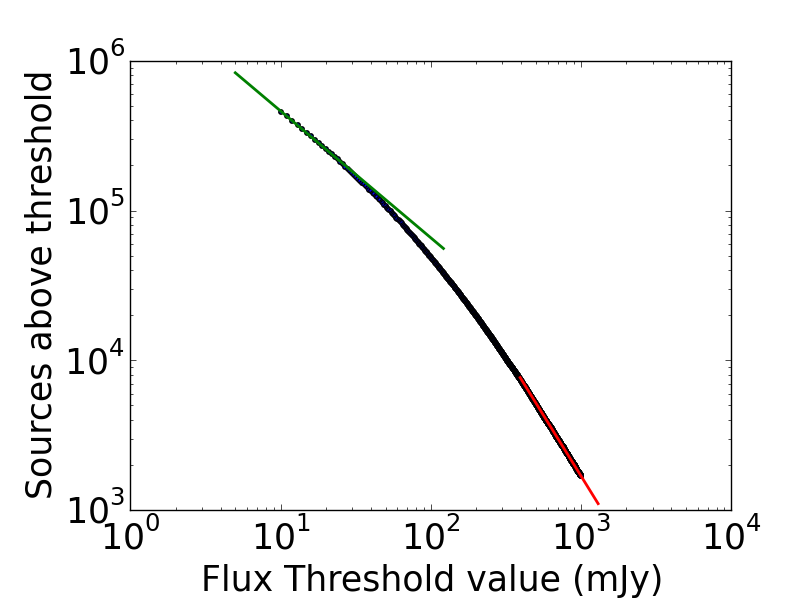} 
\caption{
The left panel is the Aitoff projection in equatorial coordinates (RA and dec) of the full NVSS catalogue. The right panel shows how the integral number count changes with the flux threshold. 
}
\label{fig:nvss}
\end{figure}

\begin{figure*}
\includegraphics[width=0.55\columnwidth]{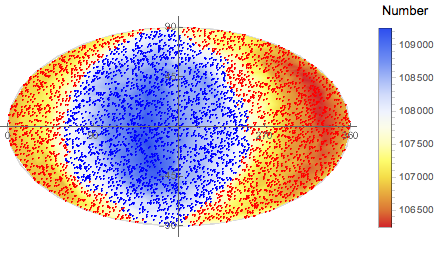} 
\includegraphics[width=0.55\columnwidth]{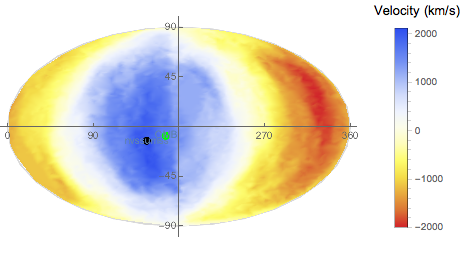}
\includegraphics[width=0.85\columnwidth]{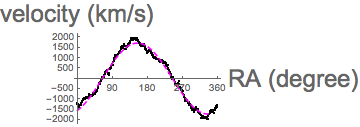}
\caption{
NVSS: The left  panel shows the directions of 5000 randomly selected hemispheres which are used for number counts. The blue dots represent hemispheres with more-than-average number of galaxies, while red dots indicate those with less-than-average number of galaxies. The middle panel is a density plot of the magnitude of the velocity, upon which the directions of the CMB dipole (green circle) and the inferred direction for the radio sources (black circle) are marked. The right panel is the plot of the derived velocities (black dots) against the RA, which is fitted well by a cosine curve (magenta).
}
\label{fig:nvss-results}
\end{figure*}

\begin{figure*}
\includegraphics[width=0.65\columnwidth]{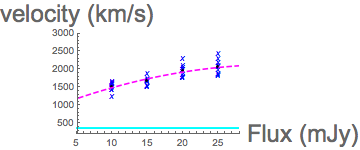}
\includegraphics[width=0.65\columnwidth]{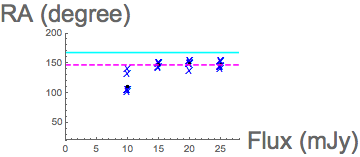}
\includegraphics[width=0.65\columnwidth]{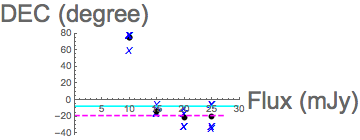}
\caption{
NVSS: The magnitude and direction of the velocity inferred from the radio dipole with various Galactic and super-Galactic cuts, and also with the LRS and the 2MRS sources removed and different patching schemes (black crosses). The black circles are the mean values at each flux and the dashed line is the fit to the mean, given by the quadratic $|v|=823 + 86S - 1.34S^2$, where $S$ is the flux threshold in mJy. Dashed (magenta) lines are the least-squares fits which give a mean velocity of $1918 \pm 298$ km\,$s^{-1}$ in the direction RA=$147\degree$, dec=$-18.5\degree$.
The solid (cyan) line is the corresponding value inferred from CMB temperature dipole. 
}
\label{fig:scatter-NVSS}
\end{figure*}

The NVSS is a catalogue of radio sources at 1.4 GHz which cover the sky north of declination $-40\degree$ (the lowest source is actually at $-40.4\degree$) corresponding to 82\% of the celestial sphere \citep{nvss89}. It contains 1,773,488 sources with flux density above 2.5 mJy. However, we restrict our analysis to sources brighter than 10 mJy as the completeness of the survey at weaker flux levels is questionable \citep{overzier03}. We also restrict the sources to be less bright than 1000 mJy as a relatively small number of very powerful sources can introduce large statistical fluctuations. 

We further cut out the catalogue above dec=$40\degree$ in order to restore the symmetry necessary for extracting the dipole \citep{singal11}. The catalogue is shown in the left panel of Fig.~\ref{fig:nvss}. The right panel shows how the integral number count changes with the threshold flux, which we use to determine the power-law index $x$ in eq.~(\ref{eq:ellisbaldwin}) finding it to be  0.894 and 1.68 at the low and high end. We cut the Galactic plane by $\pm 5\degree$, and then by $\pm 10\degree$, and study the resulting catalogues adopting flux thresholds of 10,15, 20 and 25 mJy.

The results of our hemispheric asymmetry analysis are shown in Tables \ref{tab:roya-nvss} and \ref{tab:rameez-nvss} and in Fig.~\ref{fig:nvss-results}. The results in the figure were obtained by cutting out the NVSS sources above dec$=40\degree$ to match the missing southern part below dec$=-40\degree$, and the Galactic plane was also cut at $\pm10\degree$. The dipole direction in all cases is close to that of the CMB dipole but its magnitude is on average about 4 times larger than that for the CMB (see Fig.~\ref{fig:scatter-NVSS}).

\section{Sydney University Molonglo Sky Survey (SUMSS)}
\label{sec:sumss}

\begin{figure}
\includegraphics[width=0.43\columnwidth]{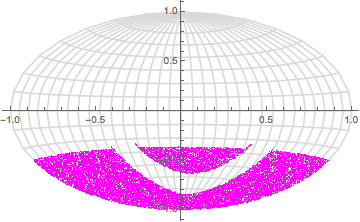} 
\includegraphics[width=0.49\columnwidth]{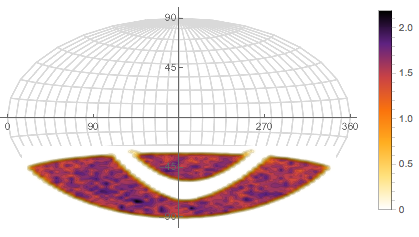} 
\caption{ 
An Aitoff projection of the SUMSS catalogue (left) and its density plot (right).
}
\label{fig:sumss}
\end{figure}

SUMSS is a radio imaging survey of the sky south of declination $-30 \degree$ carried out with the Molonglo Observatory Synthesis Telescope (MOST) operating at 843 MHz \citep{sumss03}. SUMSS contains 211,050 radio sources 
and is similar in sensitivity and resolution to the NVSS. Analysis of the catalogue shows that it is highly uniform and complete to 8 mJy at dec~$\le -50 \degree$, and to 18 mJy at $-30 \degree \ge$ dec~$> -50 \degree$. An Aitoff plot is shown in Fig.~\ref{fig:sumss}.

\begin{figure*}
\includegraphics[width=0.65\columnwidth]{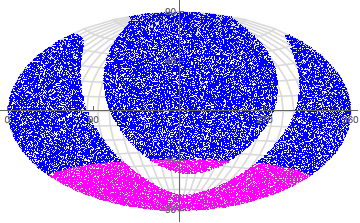} 
\includegraphics[width=0.65\columnwidth]{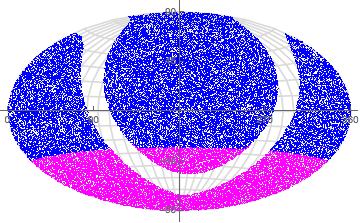} 
\includegraphics[width=0.65\columnwidth]{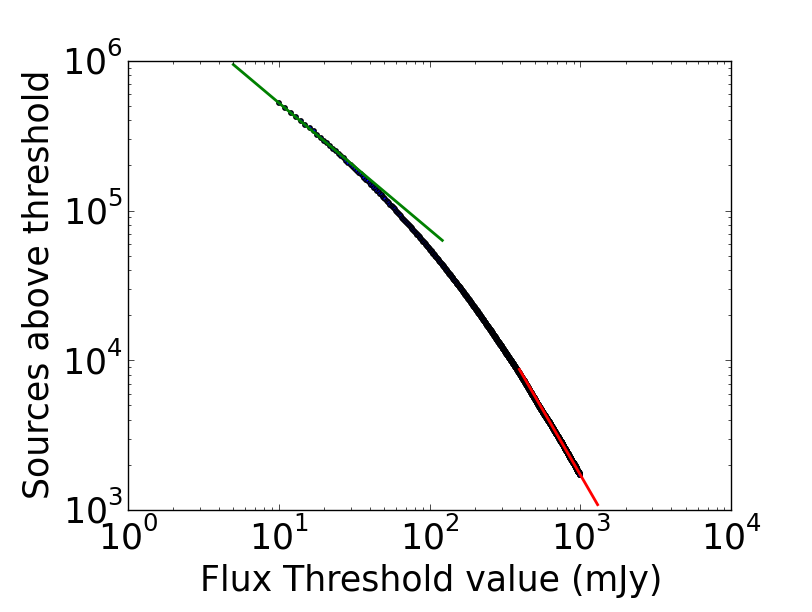} 
\caption{
NVSUMSS: The left panel shows SUMSS cut at dec=$-40\degree$ and joined with NVSS, while the middle panel shows NVSS cut at dec=$-30\degree$ and joined with SUMSS. The right panel shows the integral number of sources in NVSUMSS above a flux threshold. Also shown is the fit of eq.~(\ref{eq:ellisbaldwin}) at low and high flux which yields $x$ of 0.9 and 1.7 respectively
}
\label{fig:nvsumss}
\end{figure*}

\begin{figure*}
\includegraphics[width=0.55\columnwidth]{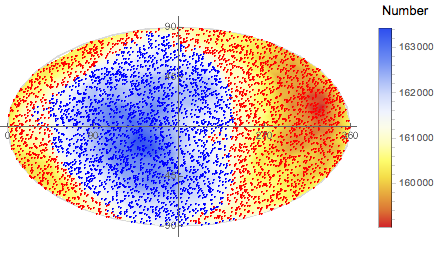} 
\includegraphics[width=0.55\columnwidth]{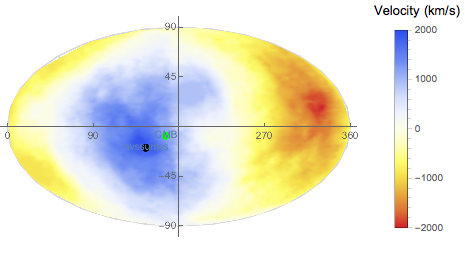}
\includegraphics[width=0.85\columnwidth]{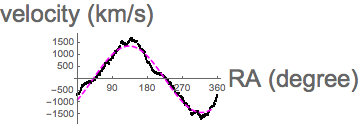}

\caption{
NVSUMSS catalogue (20--1000 mJy):
The NVSS catalogue merged with SUMSS (cut at dec=$-40\degree$) and the Galactic plane cut at latitude of $\pm 10\degree$ off NVSS to match the cut of SUMSS. The left panel shows number counts in 5000 randomly oriented hemispheres with an average of 288,218 galaxies per hemisphere and maximum difference of 5672 between the most and least populated hemispheres. The blue dots represent  hemispheres which contain more-than-average number of galaxies, while red dots indicate those with less-than-average number of galaxies. The middle panel is a density plot of the magnitude of the inferred velocity, upon which the directions of the CMB dipole (green circle) and the inferred direction for the radio sources (black circle) are marked. The right panel is the plot of inferred velocities (black dots) against RA, which is well fitted by a cosine (magenta) curve. 
}
\label{fig:nvsumss-results}
\end{figure*}

For a robust dipole analysis  a full-sky catalogue is required, so we combine NVSS and SUMSS. There are different ways of doing so which can potentially yield different results. This is discussed in the next section.

\section{Full sky catalogue: NVSS patched with SUMSS (NVSUMSS)}
\label{sec:nvsumss}

We fill in the missing sky in NVSS by combining it with its southern counterpart SUMSS. To address overlapping regions, we generate the composite catalogue in two ways. In the first, SUMSS is cut at dec=$-40\degree$ and joined to NVSS, while in the second NVSS is cut at dec=$-30\degree$ and joined to SUMSS, as shown in Fig.~\ref{fig:nvsumss}. The right panel of Fig.~\ref{fig:nvsumss} shows how the integral number count changes with the flux threshold which we use to determine the power $x$ in eq.~(\ref{eq:ellisbaldwin}).  For NVSUMSS we find it to be 0.90 (cf. 0.894 for NVSS) and 1.7 (cf. 1.68 for NVSS) at the low and high ends. To correct for the frequency difference between the SUMSS and NVSS surveys, the SUMSS fluxes are scaled down by a factor of  $(843/1400)^{-\alpha} \simeq 1.46$, where $\alpha \simeq 0.75$ is the median spectral index of the sources common to NVSS and SUMSS \citep{sumss03}.

\begin{figure*}
\includegraphics[width=0.55\columnwidth]{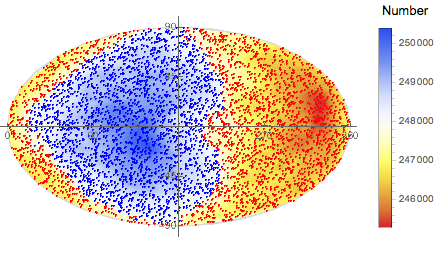} 
\includegraphics[width=0.55\columnwidth]{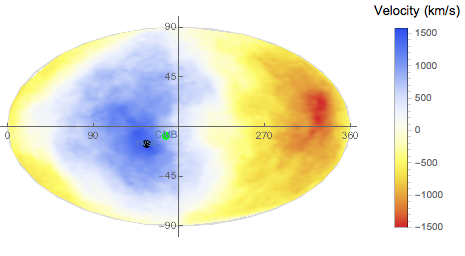}
\includegraphics[width=0.85\columnwidth]{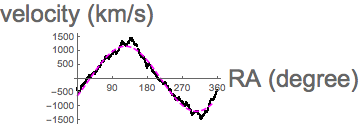}
\caption{
NVSUMSS with the Galactic plane filled with a random distribution having the same number density as the rest of the catalogue. The flux range is 15--1000 mJy and the details of the catalogues used in this figures are given in Table \ref{tab:nvsumss-GPfilled}. The left panel shows the directions of hemispheres with blue corresponding to those having average count greater than mean and red to those with average count less than the mean. The middle panel is the magnitude of the velocity inferred from the number-count hemispherical asymmetry. The right panel is obtained by regularly rotating hemispheres in the RA direction at a $1\degree$ interval. 
}
\label{fig:nvsumss-GPfilled}
\end{figure*}

Furthermore, NVSS and SUMSS have different source densities with SUMSS being denser. This can be dealt with in different ways. In the first, we randomly select galaxies from the SUMSS catalogue with declination below $-40\degree$ (or below $-30\degree$) to match the same number as NVSS above dec=$40\degree$ (or $30\degree$). In the second approach we evaluate the source density of NVSS in the full catalogue, and also that of SUMSS, and then pick randomly from SUMSS such that its density matches that of NVSS. Thirdly, when evaluating the 3-dimensional linear estimator, we weight SUMSS sources by a factor corresponding to the ratio between the number of NVSS sources above dec=$40\degree$ (or $30\degree$) and SUMSS sources below dec=$-40\degree$ (or $-30\degree$). Although in the patching scheme we run the risk of smoothing out any dipole lying in the cap regions, as we shall show these different ways of producing the composite catalogue in fact yield rather similar results. The radio dipole still implies a large velocity in a direction close to that of the CMB dipole, as shown in Tables \ref{tab:roya-nvsumss} and \ref{tab:rameez-nvsumss}, and in Figs.~\ref{fig:nvsumss-results} and \ref{fig:nvsumss-GPfilled}.

To ensure that the Galactic plane is not producing spurious effects on the inferred radio dipole, we also fill the Galactic plane with a random distribution of sources having the same surface density as the NVSUMSS catalogue. The results are preseneted in Table \ref{tab:nvsumss-GPfilled} and Figure \ref{fig:nvsumss-GPfilled} shows that they do not change significantly.

\section{Dealing with local anisotropy}
\label{sec:local}

Any clustering of nearby sources can invalidate  our approach, as both aberration and Doppler boosting analysis rely on the source distribution being isotropic so the only anisotropy is due to the observer's motion. Since the NVSS and SUMSS sources are mainly at redshift $z \sim 1$ \citep{dezotti10,ho08} local anisotropies must be minimal, nevertheless we guard against them carefully as follows. 

First we cross-correlate our composite NVSUMSS catalogue with the catalogue of nearby radio sources \citep{falcke12}. However, the latter only contains bright radio-galaxies. In order to make sure that our composite catalogue is cleaned of \emph{all} local sources, we go even further. We identify all local superclusters up to the Shapley concentration and remove them from NVSUMSS. Next we remove the super-Galactic plane in which most dominant local structures lie. The final removal of all local sources is achieved by cross-correlating NVSUMSS with the 2MRS catalogue. We thus ensure as best as we can that \emph{no} nearby sources are left in our composite catalogue.

\subsection{Cross correlation of NVSUMSS with LRS}
\label{sec:LRS}

\begin{figure*}
\includegraphics[width=0.53\columnwidth]{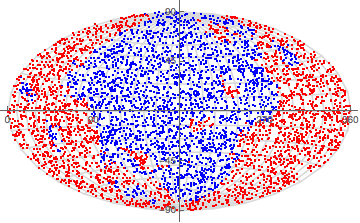} 
\includegraphics[width=0.58\columnwidth]{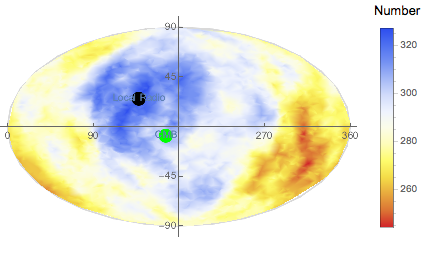} 
\includegraphics[width=0.78\columnwidth]{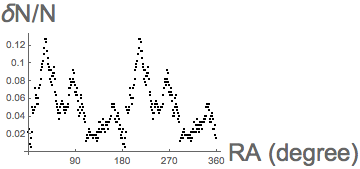} 
\caption{ 
The left panel shows the directions of 5000 randomly placed hemispheres in the LRS catalogue of 575 radio galaxies in the local universe, with blue corresponding to those having average count greater than the mean (of 288) and red to those with  average count less than the mean. The middle panel shows that a meaningful dipole cannot be seen in such a sparse and inhomogeneous sample. This is illustrated further in the right panel which plots the fractional number count difference between the forward and backward  hemispheres versus the RA of the chosen direction.
}
\label{fig:LRS}
\end{figure*}

The radio sources from the NVSS and SUMSS surveys have been matched to 2MRS galaxies using an image-level algorithm that properly treats the extended structure of radio sources \citep{falcke12}. These bright sources constitute the all-sky LRS catalogue containing 575 radio-emitting galaxies with flux greater than 213 mJy at 1.4 GHz. However this means that  nearby sources with flux less than 213 mJy are \emph{not} identified and we  shall address this in the following subsections. Here, we cross-correlate the LRS catalogue with our composite NVSUMSS catalogue and remove  zones of increasingly larger area, starting at $1/10\degree$ then $1/4\degree$ then $1/2\degree$, around the LRS sources. The results of the cross-correlation are presented in Table \ref{tab:LRS}. The dipole of the LRS, using hemispherical number count, is shown in Fig.~\ref{fig:LRS}. This shows that the contribution of LRS sources to the large-scale dipole observed in NVSS and NVSUMSS is insignificant. 

\subsection{Elimination of large superclusters up to Shapley and Ophiuchus}
 \label{ss:SC}
 
The origin of the CMB dipole is supposedly due to the gravitational attraction of nearby inhomogeneities. A large part of it can indeed be attributed to local superclusters \citep{lavaux10,colin11}. These superclusters are well observed and their position are indicated in Table \ref{tab:localsuperclusters}. (We note that in a previous study \citep{blake02} 22 local sources were removed however it is not clear what these sources were.) The results presented in Table \ref{tab:roya-nvsumss-localsuperclusters} show that the dipole velocity is not affected by these eliminations.

\subsection{Removing the super-Galactic plane}
\label{ss:SG}

\begin{figure}
\includegraphics[width=\columnwidth]{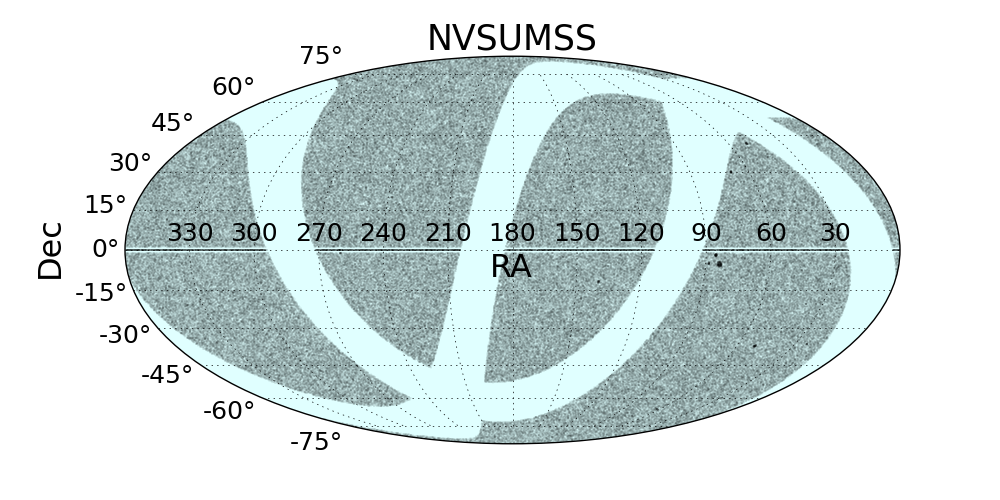}  
\caption{
NVSUMSS with Galactic and super-Galactic planes removed as strips of $\pm10\degree$ each. 
}
\label{fig:nosgp}
\end{figure}

The most important superclusters in our local Universe define the super-Galactic  plane which may contain a significant number of radio sources \citep{shaver89}. Hence one way of removing local sources is to remove this plane. Fig.~\ref{fig:nosgp} shows the NVSUMSS map with both the Galactic and super-Galactic planes cut out as $\pm 10\degree$ strips. However this makes little difference to either the magnitude or the direction of the radio dipole as seen from the results in Tables \ref{tab:2mrs-hemispheric} and \ref{tab:2mrs-3d}.

\begin{figure*}
\includegraphics[width=0.65\columnwidth]{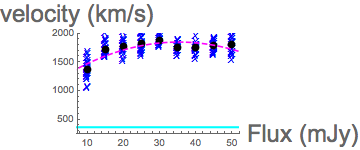}
\includegraphics[width=0.65\columnwidth]{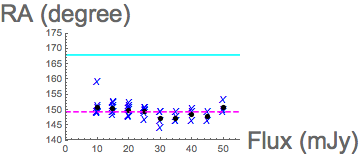}
\includegraphics[width=0.65\columnwidth]{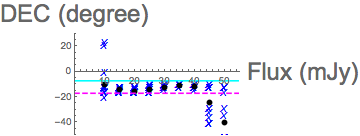}
\caption{
All results from NVSUMSS using the hemispherical number count estimator, with the Galactic and super-Galactic plane cut out and with LRS as well as 2MRS sources removed, and using different patching schemes (blue crosses). The black filled circles are the means for each flux and the dashed line is the least-squares fit to the data which yields the velocity $|v|=1121 + 41S - 0.6S^2$, where $S$ is the flux threshold in mJy. The cyan line is the value inferred from CMB temperature dipole. 
}
\label{fig:scatter-NVSUMSS-hemispheric}
\end{figure*}


\begin{figure*}
\includegraphics[width=0.65\columnwidth]{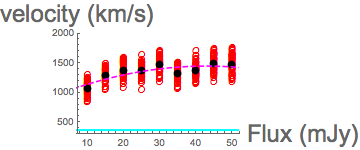}
\includegraphics[width=0.65\columnwidth]{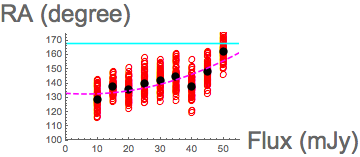}
\includegraphics[width=0.65\columnwidth]{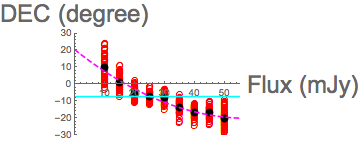}
\caption{
All results from NVSUMSS using the linear 3-dimensional estimator, with the Galactic and super-Galactic plane cut out and with LRS as well as 2MRS sources removed, and using different patching schemes (red circles). The black filled-circles are the means for each flux threshold used and the dashed line is the least-squares fit to the data which yields the velocity $|v|=926.7 + 24.4S - 0.28S^2$, where $S$ is the flux threshold in mJy. The cyan line is the value inferred from CMB temperature dipole. 
}
\label{fig:scatter-NVSUMSS-3d}
\end{figure*}
\subsection{Cross correlation of NVSUMSS with 2MRS}
\label{ss:2mrs}

There may still be local sources not lying on the super-Galactic plane, or not included in our list of prominent superclusters, or with flux less than 213 mJy. In order to be certain that \emph{all} local sources have indeed been removed, we cross-correlate NVSUMSS with the 2MRS catalogue \citep{huchra12} and remove all common objects. This is done by identifying all NVSUMMS sources that are within a chosen angular distance of 2MRS sources --- we try various values: 1, 10 and 36 arc seconds. The full set of results are shown in Figures \ref{fig:scatter-NVSUMSS-hemispheric} and \ref{fig:scatter-NVSUMSS-3d} and all details are also given in Tables \ref{tab:2mrs-hemispheric} and \ref{tab:2mrs-3d}. These demonstrate that the inferred velocity from the radio dipole remains well above that from the CMB dipole, although the directions are close.

\section{Statistical Significance}
\label{sec:ss}

To determine the statistical significance of our results, we compare them with Monte Carlo simulations using randomly generated isotropic catalogues,  on top of which aberration and Doppler boosting dipoles are added according to the null hypothesis that we are traveling through the CMB with a velocity of 369 km~s$^{-1}$ towards l=263.85$^0$, b=48.25$^0$. A sample of size $N$ is expected to contain random hemispherical asymmetries of size $\sim 1/\sqrt{N}$ on average, the direction of the  dipole being random. An aberration effect corresponding to what is expected from our velocity of 369 km~s$^{-1}$ with respect to the CMB is subsequently applied to these catalogues by shifting the angles in the direction of the CMB dipole by the expected amount. A dipole due to Doppler boosting can also be incorporated by adding (subtracting) the appropriate number of sources within angular rings oriented in (opposite to) the direction of the CMB dipole. The hemispheric asymmetry and 3-dimensional linear estimators can  subsequently be applied to the resultant random catalogues to extract the velocity. 

In 2000 mock catalogues with about 600,000 sources generated as above, the hemispheric count estimator observed velocities larger than the inferred 1600 km~s$^{-1}$ for 741 of these. The inferred dipole direction was within $15^{\degree}$ of the (simulated) CMB dipole direction for 690 of these. For the same 2000 catalogues, the 3-dimensional linear estimator found velocities higher than the inferred 1300 km~s$^{-1}$ for only 5 of the catalogues, all of which had dipole directions within $10^{\degree}$ of the (simulated) CMB dipole direction. Consequently, the statistical significance of the latter observation is at the level of $p \simeq 0.0025$, i.e $2.81\sigma$ (99.75\% confidence).

\section{The bias}
\label{sec:bias}

\begin{figure*}
\includegraphics[width=\columnwidth]{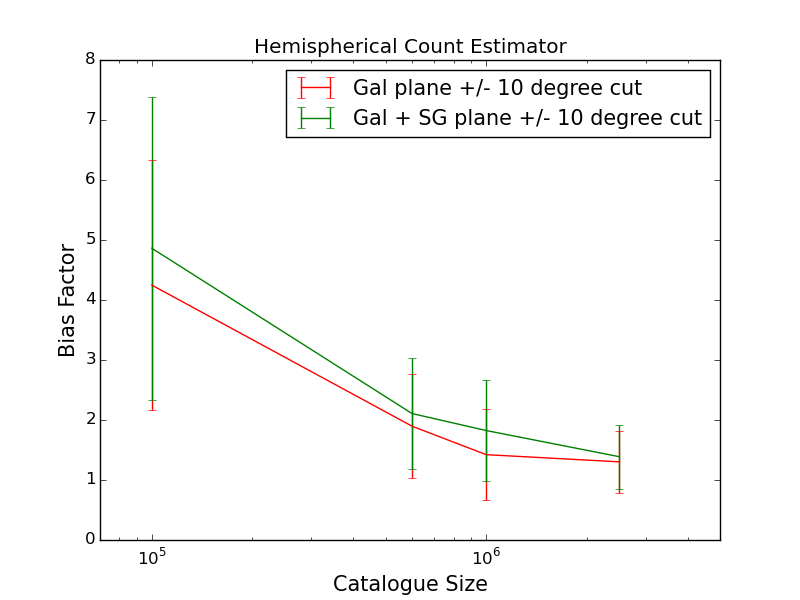}
\includegraphics[width=\columnwidth]{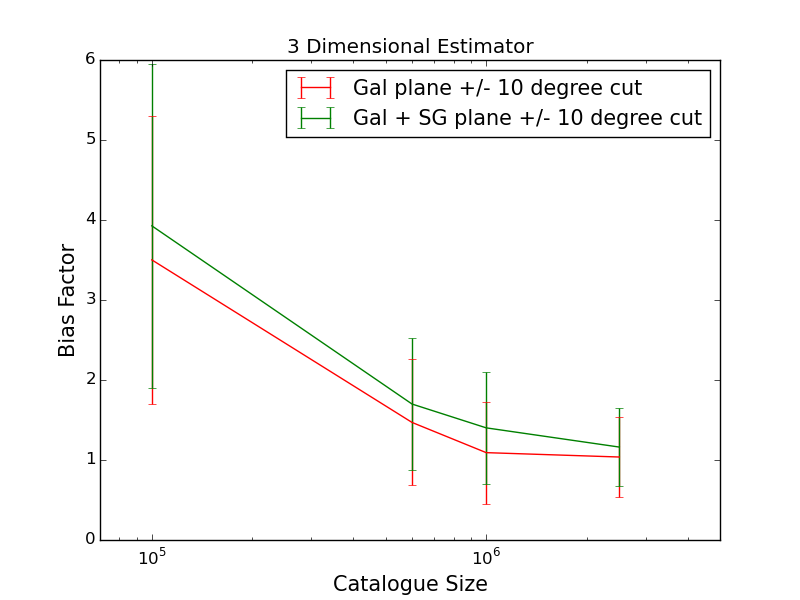}
\caption{
Bias factor versus catalogue size for the hemispheric number-counts (left) and for the 3-dimensional estimator (right).
}
\label{fig:bias}
\end{figure*}

It has been noted previously that the 3-dimensional estimator that we have used here is biased \citep{rubart13}.
Using Monte Carlo simulations we therefore evaluate: 
\begin{equation}
{\rm Bias\,\, factor}\approx {\left | {\rm True\,\, dipole + Random\,\, Dipole} \right | \over  \left | {\rm True\,\,  dipole } \right | }
\end{equation}
for both our estimators, where ``true'' refers to the CMB dipole. The bias depends on whether and by how much the Galactic plane and the super-Galactic plane are cut. We consider just two cases: the Galactic plane cut of $\pm 10 \degree$, and the Galactic as well as the super-Galactic plane cut of $\pm 10 \degree$. The results are shown in Fig.~\ref{fig:bias}. As expected the bias decreases as the sample size increases.
 The bias factor explicitly depends on the {\it true} velocity which we have assumed to be 369 km~s$^{-1}$. The bias factor for the hemispheric number-count estimator, at the catalogue size of $\sim 500,000--600,000$relevant to NVSUMSS, is about 1.9, while for the 3-dimensional estimator it is about 1.3. This explains why the inferred velocity with the former estimator is about 1600 km~s$^{-1}$ while for the latter estimator it is about 1200 km~s$^{-1}$. The agreement between them improves when the velocities are scaled down by the corresponding bias factors.

The bias does \emph{not} affect the statistical significances of our results as presented in the previous section, since it applies \emph{equally} to both the observational data as well as the null trials.

\section{Conclusion}
\label{sec:conclusion}

The velocity of the barycentre of our Solar system (or of the Local Group) in the rest frame of CMB is inferred from its temperature dipole anisotropy. An independent measurement of this velocity is needed to fully establish the kinematical origin of the dipole. Although a similar anisotropy is observed in the distribution of nearby large-scale structure out to the Shapley supercluster, they do \emph{not} show full convergence to the CMB dipole. To definitively address this question one thus needs to go further, however there is no data available from optical or infrared galaxy surveys.

Radio surveys which sample the Universe at high redshift provide valuable clues to this problem. As these scales are certainly large enough for the Universe to be considered homogeneous and isotropic (aside from negligible statistical fluctuations), the only source of deviation from isotropy would be due to the motion of the observer. Our motion should lead to a measurable over density of sources in the direction of motion through aberration and Doppler boosting effects. By measuring these effects we can deduce our velocity in the rest frame of the radio galaxies, its expected value being just  our velocity in the CMB rest frame. NVSS is a radio catalogue with less-than-complete sky coverage which has been used extensively to extract this velocity but previous studies have generally found its value to be far larger than our velocity in the CMB rest frame. We confirm this surprising result using an all-sky catalogue.

A possible reason for this discrepancy could be bias due to local structures. Of course one does not expect such anisotropies to produce a dipole \emph{larger} than the CMB dipole. However we know that our local Universe at least up to 260 Mpc ($z \simeq 0.06$) is anisotropic and if our radio catalogue is contaminated by local sources, that may invalidate the method based on the aberration and Doppler boosting effects which rely on large-scale intrinsic isotropy. We adopt various measures to remove such local sources but find that this has little effect on the inferred radio dipole.

There is a big gap in redshift between local galaxies and the radio galaxies. Forthcoming galaxy surveys will progressively fill in this gap and hopefully allow us to trace the divergence from the CMB dipole. Future radio surveys with the Square Kilometre Array will provide excellent opportunities for improving the statistical significance of the unexpectedly large radio dipole \citep{SKA15}.

\section*{Acknowledgements}

JC and RM acknowledge support from the Nordea Foundation and hospitality at the Niels Bohr International Academy. This work was supported by a DNRF Niels Bohr Professorship awarded to SS. We thank Pavel Naselsky, Andy Jackson, Dominik Schwarz and Sjoert van Velzen for helpful discussions, and especially, Prabhakar Tiwari for emphasising the need for flux rescaling in assembling the NVSUMMS flux-limited catalogue and the anonymous Referee who pointed out the need to estimate the bias in the estimators used.



\begin{table*}
\caption{
{\sf 
{Hemispheric asymmetry, NVSS (with the Northern cap removed at dec=40$\degree$ for symmetry with the missing Southern part, uses 5000 random hemispheres):}
 Galactic latitude Cut ($\pm$),  Flux threshold (mJy), Total number of sources left after cuts,
 Average numbers in each hemisphere ${\bar N}$, Maximum difference in number count of hemispheres $\Delta N$, Dipole RA (deg), Dipole dec (deg), Velocity (km~s$^{-1}$)
}}
\begin{tabular}{|c|c|c|c|c|c|c|c|c|c|c|}
& & & & & & & & & & \\
b cut & Flux cut & $N$ & ${\bar N}$ & $\Delta N$ &RA$\degree$ & dec$\degree$ & v (km/s) \\
\hline
5  & 10 & 416389 & 208180 & 4580 & 100.6 & 77.1 & 1657.5 \\
5  & 15 & 299409 & 149704 & 3762 & 142.9 & -6.1  &  1899.2 \\
5  & 20 & 233277 & 116631 & 3565 & 148.2 & -10.8 & 2294.9 \\
5 & 25 & 189995 & 95012 & 3081 & 150.6 & -35.1 & 2438.8 \\
10 & 10 & 384844 & 192391 & 4096 & 104.6 & 77.3 & 1644.8 \\
10 & 15 & 276524 & 138281 & 3220 & 148.1 & -10.9 & 1732.6 \\
10 & 20 & 215418 & 107725 & 3018 & 148.9 & -11.5 & 2119.2 \\
10 & 25 & 175383 & 87713 & 2690 & 143.9 & - 5.9 & 2296.4 \\
&&&\\
\hline
\label{tab:roya-nvss}
\end{tabular}
\end{table*}


\begin{table*}
\caption{
{\sf 
{Hemispheric asymmetry, NVSS (with the Northern cap removed at dec=40$\degree$ for symmetry with missing Southern part. Using HEALPix grid of Nside=32):}
super-Galactic latitude cut ($\pm$), Galactic latitude Cut ($\pm$), Flux threshold (mJy), Total number of sources left after cuts, Dipole dec (deg), Dipole RA (deg), Dipole value, Galactic b (deg), Galactic l (deg), Velocity (km~s$^{-1}$)
}}
\begin{tabular}{|c|c|c|c|c|c|c|c|c|c|}
& & & & & & & & & \\
SGB cut & b cut & Flux cut & $N$ & dec$\degree$ & RA$\degree$ & D & b$\degree$ & l$\degree$ & v (km/s) \\ 
\hline
0&5&10&413254&76.8&105.0&0.012&26.8&137.7&1502\\
0&5&15&297806&-17.0&150.5&0.013&29.7&255.0&1668\\
0&5&20&232256&-32.8&153.3&0.016&19.2&268.4&2054\\
0&5&25&189357&-32.8&153.3&0.017&19.2&268.4&2183\\
0&10&10&381951&76.8&105.0&0.011&26.8&137.7&1447\\
0&10&15&275168&-15.7&149.1&0.012&29.7&252.9&1530\\
0&10&20&214643&-17.0&147.7&0.014&27.8&252.8&1859\\
0&10&25&174881&-6.0&143.4&0.015&31.9&240.1&1992\\
5&5&10&384647&76.8&105.0&0.012&26.8&137.7&1619\\
5&5&15&277131&-17.0&150.5&0.013&29.7&255.0&1682\\
5&5&20&216046&-32.8&153.3&0.015&19.2&268.4&2002\\
5&5&25&176136&-32.8&153.3&0.016&19.2&268.4&2105\\
5&10&10&353344&76.8&105.0&0.012&26.8&137.7&1569\\
5&10&15&254493&-15.7&149.1&0.012&29.7&252.9&1535\\
5&10&20&198433&-17.0&147.7&0.014&27.8&252.8&1785\\
5&10&25&161660&-6.0&143.4&0.015&31.9&240.1&1891\\
10&5&10&356540&58.9&139.3&0.012&41.6&156.6&1532\\
10&5&15&256877&-17.0&150.5&0.013&29.7&255.0&1703\\
10&5&20&200321&-32.8&153.3&0.015&19.2&268.4&2000\\
10&5&25&163267&-32.8&153.3&0.016&19.2&268.4&2081\\
10&10&10&325237&76.8&105.0&0.011&26.8&137.7&1432\\
10&10&15&234239&-15.7&149.1&0.012&29.7&252.9&1546\\
10&10&20&182708&-17.0&147.7&0.014&27.8&252.8&1765\\
10&10&25&148791&-6.0&143.4&0.014&31.9&240.1&1846\\
&&&\\
\hline
\label{tab:rameez-nvss}
\end{tabular}
\end{table*}



\begin{table*}
\caption{
{\sf 
{Hemispheric asymmetry, NVSUMSS (using 5000 random hemispheres):}
 Galactic latitude cut ($\pm$), NVSS-SUMSS patch declination, Flux threshold (mJy), Total number of sources left after cuts,
 Average numbers in each hemisphere ${\bar N}$, Maximum difference in number counts of hemsipheres $\Delta N$, Dipole dec (deg), Dipole RA (deg), Velocity (km~s$^{-1}$)
}}
\begin{tabular}{|c|c|c|c|c|c|c|c|c|}
& & & & & & & & \\
b cut & patch cut & Flux cut & $N$ & ${\bar N}$ & $\Delta N$ & RA$\degree$ & Dec$\degree$ & v (km/s) \\
\hline
10 &-40. & 10 & 576461 & 288212 & 5880 & 151.3 & -16.5 & 1570.9 \\
10 &-40. & 15 & 413838 & 206932 & 5165 & 148.2 & -13.1 & 1861.6 \\
10 &-40. & 20 & 322452 & 161229 & 4422 & 148.0 & -16.8 & 2056.1 \\
10 &-40. & 25 & 262617 & 131297 & 3538 & 149.8 & -12.9 & 1990.5 \\
10 &-30. & 10 & 576617 & 288311 & 6468 & 148.8 & -1.3 &  1673.0 \\
10 &-30. & 15 & 413564 & 206768 & 5456 & 152.6 & -17.4 & 1999.9 \\
10 &-30. & 20 & 322074 & 161020 & 4304 & 151.3 & -15.0 & 2017.6 \\
10 &-30. & 25 & 262332 & 131149 & 3510 & 149.5 & -13.3 & 2005.9 \\
&&&\\
\hline
\label{tab:roya-nvsumss}
\end{tabular}
\end{table*}


\begin{table*}
\caption{
{\sf 
{Hemispheric asymmetry, NVSUMSS (using HEALPix grid of Nside=32):}
super-Galactic latitude cut ($\pm$), Galactic latitude Cut ($\pm$), NVSS-SUMSS patch declination, Flux threshold (mJy), Total number of sources left after cuts, Dipole dec (deg), Dipole RA (deg), Dipole value, Galactic b (deg), Galactic l (deg), Velocity (km~s$^{-1}$)
}}
\begin{tabular}{|c|c|c|c|c|c|c|c|c|c|c|}
& & & & & & & & & & \\
SGB cut & b cut & patch cut & Flux cut & $N$ &dec$\degree$ & RA$\degree$ & D & b & l & v (km/s) \\
\hline
0&10&-40&10&571017&-13.2&149.1&0.0084&31.4&250.8&1094\\
0&10&-40&15&410966&-13.2&149.1&0.0111&31.4&250.8&1436\\
0&10&-40&20&320577&-17.0&150.5&0.0117&29.7&255.0&1520\\
0&10&-40&25&261293&-17.0&150.5&0.0128&29.7&255.0&1656\\
0&10&-30&10&571982&-13.2&149.1&0.0076&31.4&250.8&987\\
0&10&-30&15&411258&-13.2&149.1&0.0109&31.4&250.8&1409\\
0&10&-30&20&320651&-17.0&150.5&0.0118&29.7&255.0&1530\\
0&10&-30&25&261363&-17.0&150.5&0.0113&29.7&255.0&1461\\
5&10&-40&10&517269&-13.2&149.1&0.0090&31.4&250.8&1169\\
5&10&-40&15&372124&-17.0&150.5&0.0113&29.7&255.0&1463\\
5&10&-40&20&290194&-15.7&151.9&0.0116&31.5&255.1&1507\\
5&10&-40&25&236448&-10.8&146.3&0.0124&31.0&246.5&1609\\
5&10&-30&10&518021&-13.2&149.1&0.0088&31.4&250.8&1143\\
5&10&-30&15&372370&-17.0&150.5&0.0108&29.7&255.0&1397\\
5&10&-30&20&290249&-17.0&147.7&0.0111&27.8&252.8&1438\\
5&10&-30&25&236574&-13.2&149.1&0.0115&31.4&250.8&1486\\
10&10&-40&10&464161&-13.2&149.1&0.0100&31.4&250.8&1295\\
10&10&-40&15&334073&-13.2&149.1&0.0129&31.4&250.8&1673\\
10&10&-40&20&260580&-14.5&150.5&0.0118&31.5&253.0&1529\\
10&10&-40&25&212239&-10.8&146.3&0.0120&31.0&246.5&1551\\
10&10&-30&10&464926&-13.2&149.1&0.0098&31.4&250.8&1265\\
10&10&-30&15&334335&-13.2&149.1&0.0127&31.4&250.8&1646\\
10&10&-30&20&260539&-13.2&149.1&0.0118&31.4&250.8&1530\\
10&10&-30&25&212219&-13.2&149.1&0.0116&31.4&250.8&1503\\
&&&\\
\hline
\label{tab:rameez-nvsumss}
\end{tabular}
\end{table*}

\begin{table*}
\caption{
{\sf 
{Hemispheric asymmetry, NVSUMSS (with Galactic plane filled with a random distribution; uses 5000 random hemispheres):}
 Galactic latitude cut ($\pm$), NVSS-SUMSS patch declination, Flux threshold (mJy), Total number of sources left after cuts,
 Average numbers in each hemisphere ${\bar N}$, Maximum difference in number counts of hemispheres $\Delta N$, Dipole RA (deg), Dipole dec (deg), Velocity (km~s$^{-1}$)
}}
\begin{tabular}{|c|c|c|c|c|c|c|c|c|}
& & & & & & & & \\
b cut & patch cut & Flux cut & $N$ & ${\bar N}$ & $\Delta N$ & RA$\degree$ & dec$\degree$ & v (km/s) \\
\hline
10 &-40 & 10 & 689995 & 345007 & 5743 & 149.0 & -14.3 & 1306.3 \\
10 &-40 & 15 & 495573 & 247804 & 5164 & 149.2 & -13.8 & 1590.6 \\
10 &-40 & 20 & 386549 & 193283 & 4428 & 150.0 & -17.3 & 1722.6 \\
10 &-40 & 25 & 314213 & 157107 & 4200 & 148.8 & -13.3 & 2003.6 \\
10 &-30 & 10 & 690314 & 345146 & 6954 & 159.0 & -20.9 & 1512.8 \\
10 &-30 & 15 & 495366 & 247685 & 5432 & 152.3 & -16.2 & 1647.3 \\
10 &-30 & 20 & 385862 & 192957 & 4371 & 152.3 & -16.2 & 1711.2 \\
10 &-30 & 25 & 313849 & 156947 & 4193 & 149.6 & -14.5 & 2006.9 \\
&&&\\
\hline
\label{tab:nvsumss-GPfilled}
\end{tabular}
\end{table*}


\begin{table*}
\caption{
{\sf 
{Hemispheric asymmetry, NVSUMSS (using 5000 random hemispheres):}
Area in degree$^2$ around sources in LRS which is removed from NVSUMSS, Galactic latitude cut ($\pm$), NVSS-SUMSS patch declination, Flux threshold (mJy), Total number of sources left after cuts, Average numbers in each hemisphere ${\bar N}$, Maximum difference in number counts of hemsipheres $\Delta N$, Dipole RA (deg), Dipole dec (deg), Velocity (km~s$^{-1}$).}}
\begin{tabular}{|c|c|c|c|c|c|c|c|c|c|c|c|}
& & & & & & & & & & &\\
Area cut & b cut & patch cut & flux cut & $N$ & ${\bar N}$ & $\Delta N$ & RA$\degree$ & dec$\degree$ & v (km/s) \\
\hline
${1/10\degree}^2$ & 10 &-40. & 15 & 411828 & 205918 & 5275 & 150.2 & -13.8 & 1922.4 \\
${1/4\degree}^2$  & 10 &-40. & 15 & 410745 & 205369 & 5072 & 150.6 & -17.4 & 1842.4 \\
${1/2\degree}^2$  & 10 &-40. & 15 & 409027 & 204511 & 4745 & 152.5 & -17.2 & 1768.7 \\
&&&\\
\hline
\label{tab:LRS}
\end{tabular}
\end{table*}


\begin{table*}
\caption{
{\sf 
{Hemispheric asymmetry, NVSUMSS (using 5000 random hemispheres):} Area in degree$^2$ around the 8 local superclusters (see Table \ref{tab:localsuperclusters}), Galactic latitude cut ($\pm$), NVSS-SUMSS patch declination, Flux threshold (mJy), Total number of sources left after cuts, Average numbers in each hemisphere ${\bar N}$, Maximum difference in number counts of hemsipheres $\Delta N$, Dipole RA (deg), Dipole dec (deg), Velocity (km~s$^{-1}$). In total about 5000 to 6000 sources are removed.
}}
\begin{tabular}{|c|c|c|c|c|c|c|c|c|c|}
& & & & & & & & &\\
Area cut & b cut & patch cut & flux cut & $N$ & ${\bar N}$ & $\Delta N$ & RA$\degree$ & dec$\degree$ & v (km/s) \\
\hline
${1/10\degree}^2$ & 10 &-40 & 15 & 409025 & 204477 & 4709 & 151.4 & -16.5 & 1788.5 \\
${1/4\degree}^2$  & 10 &-40 & 15 & 409011 & 204488 & 4885 & 151.4 & -16.4 & 1805.5 \\
${1/2\degree}^2$  & 10 &-40 & 15 & 408989 & 204492 & 4683 & 146.8 & -12.8 & 1703.6 \\
&&&\\
\hline
\label{tab:roya-nvsumss-localsuperclusters}
\end{tabular}
\end{table*}


\begin{table*}
\caption{
{\sf Hemispheric asymmetry, NVSUMSS cross-correlated with 2MRS to remove local sources (using HEALPix grid of Nside=32)}.
Only a sample of results are shown (full table is available on the online version).
{\small (1) Angular tolerance within which 2MRS sources are looked for, (2) Number of sources removed because of 2MRS overlap, (3) super-Galactic latitude cut ($\pm$), (4) Galactic Latitude cut ($\pm$), (5) NVSS-SUMSS patch declination, (6) Flux threshold (mJy), (7) Total number of sources left after cuts, (8) Dipole dec (deg), (9) Dipole RA (deg), (10) Dipole value, (11) Galactic b (deg), (12) Galactic l (deg), (13) Velocity (km~s$^{-1}$)}}
\begin{tabular}{|c|c|c|c|c|c|c|c|c|c|c|c|c|}
& & & & & & & & & & &\\
(1) & (2) & (3) & (4) & (5) & (6) & (7) & (8) & (9) & (10) & (11) & (12) & (13) \\
\hline
10"&3701&10&10&-40&10&460484&-13.2&149.1&0.009&31.4&250.8&1368\\
10"&2510&10&10&-40&15&331552&-13.2&149.1&0.011&31.4&250.8&1792\\
10"&1847&10&10&-40&20&258813&-13.2&149.1&0.012&31.4&250.8&1873\\
10"&1470&10&10&-40&25&210756&-13.2&149.1&0.011&31.4&250.8&1742\\
10"&1228&10&10&-40&30&177191&-13.2&149.1&0.012&31.4&250.8&1898\\
10"&1036&10&10&-40&35&152561&-10.8&146.3&0.012&31.0&246.5&1809\\
10"&897&10&10&-40&40&133018&-10.8&146.3&0.011&31.0&246.5&1784\\
10"&788&10&10&-40&45&117685&-13.2&149.1&0.012&31.4&250.8&1811\\
10"&710&10&10&-40&50&105184&-52.8&174.6&0.013&8.5&291.9&2025\\
10"&3662&10&10&-30&10&461202&-13.2&149.1&0.009&31.4&250.8&1484\\
10"&2499&10&10&-30&15&331651&-13.2&149.1&0.010&31.4&250.8&1629\\
10"&1846&10&10&-30&20&258767&-13.2&149.1&0.011&31.4&250.8&1740\\
10"&1456&10&10&-30&25&210719&-12.0&144.8&0.011&29.2&246.6&1679\\
10"&1209&10&10&-30&30&177340&-13.2&149.1&0.011&31.4&250.8&1749\\
10"&1016&10&10&-30&35&152334&-10.8&146.3&0.010&31.0&246.5&1597\\
10"&877&10&10&-30&40&133015&-10.8&146.3&0.011&31.0&246.5&1795\\
10"&776&10&10&-30&45&117626&-13.2&149.1&0.012&31.4&250.8&1906\\
10"&699&10&10&-30&50&105089&-52.8&174.6&0.012&8.5&291.9&1829\\
&&&&&&&&&&&\\
\hline
\label{tab:2mrs-hemispheric}
\end{tabular}
\end{table*}

\begin{table*}
\caption{
{\sf 3-dimensional linear estimator, NVSUMSS cross-correlated with 2MRS to remove local sources}.
Only a sample of results are shown (full table is available on the online version).
{\small (1) Angular tolerance within which 2MRS sources are looked for, (2) Number of sources removed because of 2MRS overlap, (3) super-Galactic latitude cut ($\pm$), (4) Galactic latitude cut ($\pm$), (5) NVSS-SUMSS patch declination, (6) Flux threshold (mJy), (7) Total number of sources left after cuts, (8) Dipole dec (deg), (9) Dipole RA (deg), (10) Dipole value, (11) Galactic b (deg), (12) Galactic l (deg), (13) Velocity (km~s$^{-1}$)}}
\begin{tabular}{|c|c|c|c|c|c|c|c|c|c|c|c|c|}
& & & & & & & & & & &\\
(1) & (2) & (3) & (4) & (5) & (6) & (7) & (8) & (9) & (10) & (11) & (12) & (13) \\
\hline
10"&3654&10&10&-40&10&457184&14.1&128.4&2468&28.9&211.2&1265\\
10"&2465&10&10&-40&15&329034&5.9&128.2&1755&25.2&219.4&1250\\
10"&1807&10&10&-40&20&256430&-1.7&126.3&1453&19.9&225.7&1328\\
10"&1423&10&10&-40&25&208861&-4.3&127.6&1156&19.7&228.8&1297\\
10"&1196&10&10&-40&30&175595&-4.4&134.3&1058&25.4&232.7&1412\\
10"&1006&10&10&-40&35&150883&-11.5&136.8&840&23.3&240.7&1304\\
10"&870&10&10&-40&40&131662&-15.6&129.6&859&15.3&239.8&1529\\
10"&767&10&10&-40&45&116504&-15.3&143.0&751&25.7&248.0&1512\\
10"&693&10&10&-40&50&104075&-23.1&156.3&667&28.6&264.3&1502\\
10"&3602&10&10&-30&10&455766&1.0&126.2&2417&21.1&223.2&1243\\
10"&2424&10&10&-30&15&327749&-0.4&125.8&1753&20.1&224.3&1253\\
10"&1769&10&10&-30&20&255327&-4.9&123.9&1394&16.2&227.4&1280\\
10"&1392&10&10&-30&25&208009&-8.5&122.9&1078&13.5&230.0&1215\\
10"&1171&10&10&-30&30&174886&-6.0&130.3&1003&21.1&231.8&1344\\
10"&989&10&10&-30&35&150122&-9.8&133.0&782&21.3&236.7&1221\\
10"&858&10&10&-30&40&131043&-12.1&128.2&839&16.1&236.1&1501\\
10"&753&10&10&-30&45&115919&-12.5&139.6&712&24.9&243.4&1440\\
10"&677&10&10&-30&50&103530&-17.0&152.8&625&31.2&256.9&1415\\
\hline
\label{tab:2mrs-3d}
\end{tabular}
\end{table*}

\end{document}